\documentstyle[12pt]{ioplppt}

\begin{document}

\jl{3}

\newcommand{\ie}{{\it i.\ e.\ }}
\newcommand{\etc}{{\it etc.\ }}
\newcommand{\eg}{{\it e.\ g.\ }}
\newcommand{\be}{\begin{eqnarray}}
\newcommand{\en}{\end{eqnarray}}
\newcommand{\no}{\nonumber}
\newcommand{\hc}{{\it h.\ c.\ }}

\title{XPS and NMR as Complementary Probes of Pseudogaps 
and Spin--charge Separation.}[XPS\ 'v'\ NMR]

\author{Nic\ Shannon}

\address{Department of Physics, University of Warwick,
Coventry CV4 7AL, UK.}

\date{\today}

\begin{abstract}
The possibility that strongly correlated many--electron systems 
may exhibit spin--charge separation has generated great excitement,
particularly in the light of recent experiments on low dimensional 
conductors and high temperature superconductors.   However, finding 
experimental support for this hypothesis has been made difficult by 
the fact that most commonly used probes couple simultaneously to 
spin {\it and} charge excitations.   We argue that core hole
photoemission (XPS)/nuclear magnetic resonance (NMR) couple
independently and in exactly comparable ways to the local charge/spin 
susceptibilities of the system being measured.  The explicit comparison
of XPS and NMR data, particularly in systems which exhibit a 
pseudogap, may therefore yield fresh evidence for the existence (or 
non--existence) of spin--charge separation.   
Application of these ideas to the normal 
state of high temperature superconductors is discussed, and 
the application is further illustrated in some detail for
quasi one--dimensional systems with charge density waves. 
\end{abstract}


\maketitle

\section{Introduction}

While the words ``spinon'' and ``holon'' are now firmly entrenched in the 
common vernacular of condensed matter physics, hard experimental evidence 
for the existence of low energy spin--charge separation in real materials 
has proved elusive.    {\it Pseudogaps} --- by which we mean 
behavior comparable to the opening of a gap but exhibited in the 
``normal'' state of a system with an ordered groundstate --- are on the other 
hand quite ubiquitous in 
those systems which are candidates for a spin--charge separated description. 
In this paper we will discuss what might be learned from the explicit
comparison of core hole photoemission (XPS) and nuclear magnetic 
resonance (NMR) about two classes 
of material which display pseudogap behavior and may be spin--charge 
separated --- the cuprate high temperature superconductors and 
quasi--one dimensional charge density wave systems.

The main reason for the difficulty in determining empirically whether a
given system is spin--charge separated is that most experimental probes couple to 
both spin and charge excitations.
In the case of angle resolved photoemission (ARPES), potentially one of the 
most powerful probes, a physical electron must be removed from the 
system under investigation, changing both its spin and charge quantum numbers.
In order to extract information about spin--charge separation one must 
therefore have a concrete model giving predictions which are wholly dependent 
on the existence of spin--charge separation, against which the data may be 
checked.   For this reason the search for spin--charge separation in low 
dimensional materials has often been equated with the search for hints of 
Luttinger Liquid (LL) behavior (noteably separate dispersing spin and 
charge peaks \cite{volker}) in  ARPES spectra (see \eg 
\cite{chen,jim,kobayashi,yves}).   
Ingenious alternative approaches, such as the examination of 
spin--diffusion \cite{si} in a 1D system, or the tunneling between 1D and 2D
electron liquids \cite{andy}, depend on exploiting a 
difference in the decay times or velocities for collective spin and charge 
excitations, and do not probe spin and charge separately.
We note that evidence for Luttinger Liquid behaviour has been found 
recently from transport experiments on metallic 
Carbon Nanotubes \cite{david}.

More powerful general statements could be made about systems for which 
no detailed model existed if it were possible to directly compare
two experiments which coupled independently to spin and charge excitations.   
We argue here that NMR, which has been 
used extensively in the investigation of pseudogap behavior in the 
cuprate high temperature superconductors \cite{T*}, couples to spin 
excitations {\it in exactly the same way} that XPS couples to charge 
excitations, and that the two
are therefore useful complementary probes of spin--charge separated 
behavior.   We will also discuss how the existence of a pseudogap 
may in fact make it easier to distinguish spin--charge separated and
non spin--charge separated theories.

We first briefly define spin--charge separation and 
review the relevant many body effects which come
into play in XPS and NMR.   Since the goal of this 
paper is to motivate the comparison of 
two different experiments
and not to provide a definitive treatment of
any given set of results, 
we adopt a somewhat naive picture of both XPS and NMR.
We hope in later papers to relax some
of these simplifying assumptions. 

\section{Spin--charge separation}

In atomic physics, quantum numbers may be classified by the symmetries
to which they correspond, and broken up into groups according to the 
type of excitation (transition) considered.   The assertion that
a many body system is spin--charge separated is the claim that, 
at least at low energies, its energy levels may be classified by two 
{\it independent} sets of quantum numbers, one for spin and one for 
charge excitations.  In terms of the Hamiltonian, this means 
\be
\label{eqn:defn}
{ H} &=& { H_{\rho}} + { H_{\sigma}} 
   \qquad [{ H_{\rho}}, { H_{\sigma}} ] = 0
\en
where ${ H_{\rho}}$ acts on the (Hilbert) space for charge and 
${ H_{\sigma}}$ on the space for spin excitations.   The condition
\eref{eqn:defn} may not be strictly realized by microscopic model 
Hamiltonians relevant to experiment, but it remains a useful concept as 
long as coupling between spin and charge modes is sufficiently weak in 
the low energy limit.

As a consequence of \eref{eqn:defn} the state of the system factorizes
\be
\mid \Psi_{n} \rangle &=& 
   \mid \vec{\rho} \rangle \times \mid \vec{\sigma} \rangle
\en
where $n$ is the principal quantum number of the state and
$\vec{\rho}$ and $\vec{\sigma}$ are (sets of) quantum numbers
for spin and charge excitations.

Such a factorization would of course come as no surprise in the 
real--space world of atomic physics, but the independent electron model of 
metals (and band insulators) teaches us to think in terms of Bloch states  ---
excitations of the many electron system are counted using the momentum quantum 
numbers of individual electrons, which means that quasiparticles {\it 
must} carry both the spin and charge of an electron.   Interaction changes 
this picture very little within the usual Fermi liquid 
scenario, since its low lying excitations are by assumption
electron--like quasiparticles with good momentum quantum numbers.  Many different 
soluble models of interacting electrons in one dimension (for example 
the Tomonaga--Luttinger model discussed below) {\it do} however
display spin--charge separation.

Spin--charge separation is then inherently a strong correlation effect --- 
interaction between electrons should be important enough to render them 
completely incoherent, whilst at the same time stabilizing independent 
sets of quasi--particles with 
pure spin and charge quantum numbers (``spinons'' and ``holons'').   
In one dimension this occurs
for arbitrarily weak interaction because charge confined to a line 
suffers rigorous phase space constraints.   In higher dimension 
interaction must be much stronger, or the phase space for electrons 
constrained by other considerations.   We now turn to experiment. 

\section{XPS and NMR}

\subsection{Experimental response functions.}

NMR is an indirect probe of the spin susceptibility of an electron 
liquid.   Excitations of the Nuclear lattice in a solid are damped by their 
interaction with the orbital and spin degrees of freedom of electrons.
In a metal, the most important contribution to this damping comes from 
a hyperfine interaction with itinerant electrons and may be expressed 
in terms of a nuclear relaxation time $T_1$ according to 
\be
\frac{1}{T_1T} &=& \lim_{\omega \to 0} 
   \sum_q F(q) \frac{\Im \{ \chi_{\sigma}(q,\omega) \} }{\omega}
\en
where $T$ is the temperature of the system, $\chi_{\sigma}(q,\omega)$
the dynamic spin--susceptibility of the electrons, and all relevant coupling 
constants have been absorbed into a weakly momentum dependent structure factor 
$F(q)$.

Provided that the momentum dependence of the structure factor can
be neglected ($F(q) \to F_0$, equivalent to assuming a delta function 
interaction in real space), $1/T_1T$ is directly related to the 
imaginary part of the local spin--susceptibility.  Evaluating this
as single electron--hole bubble one then finds
\be
\frac{1}{T_1T} &=& F_0 n_0^2
\en
where $n_0$ is the density of states at the Fermi surface.
The result $1/T_1T = const.$ is known as the Korringa law.

Under certain plausible assumptions \cite{citrin}, XPS measures the spectral function 
$A_c(\omega)$ of 
core electrons, which in most cases are highly localized and have 
very large binding energies.  
Core states in a metal are coupled by coulomb interaction
with the itinerant charge of the conduction band.  An infrared
divergence in the number of charge carrying excitations
of the itinerant electron liquid made by the photoemission of the core
electron powerlaw divergence at threshold in their spectral function 
\be
\label{eqn:power}
A_c(\omega) \sim \frac{1}{(\omega-\epsilon_c)^{1-\alpha}}
\en
characterized by an exponent $\alpha$.  In the limit of weak coupling 
(essentially a second order linked cluster expansion \cite{note1}), we 
find
\be
\alpha &=& \lim_{\omega \to 0} 
   \sum_q \mid V_q \mid^2 \frac{\Im \{ \chi_{\rho}(q,\omega) \} }{\omega}
\en
and for an ordinary metal, under the assumption of local 
interaction ($V_q \to V_0$), 
\be
\label{eqn:born}
\alpha &=& \mid V_0 \mid^2 n_0^2
\en
which is once again independent of temperature.   

Here we have deliberately adopted an idealized view of XPS line shapes,
which are only strictly simple power laws at threshold and 
zero temperature, 
and in practice are considerably broadened by the (temperature
independent) Auger decay of the core hole and a temperature dependent 
coupling to phonons.  
However none these mechanisms greatly affects the 
asymmetry of the line, which is still determined by $\alpha$, 
and good fits to experiment can usually be found by convoluting
the simple power law Eq.\ \eref{eqn:power} with Lorentzian (Auger 
process) and Gaussian (phonon process) lifetime envelopes to give 
the widely accepted Doniach--Sunjic lineshape \cite{doniach}.

\subsection{Relevance to spin--charge separation}

Within the picture developed above we see that XPS and NMR are 
{\it local zero frequency probes}, 
one coupling to charge and the other to spin excitations.  Under the 
assumption of delta function interaction, both the 
Nuclear relaxation time $1/T_1T$ and the core level asymmetry exponent 
$\alpha$ are related to the local spin(charge) susceptibility {\it in
exactly the same way}.  They therefore form a complementary pair of 
probes which may be used to assess the degree to which spin and charge 
excitations are linked in any given system.   

Without needing to have recourse to any specific model of the system under
investigation, we can therefore directly compare the temperature 
dependence of $\alpha$ and $1/T_1T$, either by plotting one against the
other, or by examining the ratio
\be
{ R}(T) &=& \frac{\alpha}{1/T_1T}
\en
as a function of temperature, rather as one might use the ratio of thermal 
and electrical conductivities to test the 
independent electron model of a metal (the Wiederman--Franz law).
In this case or, in a Fermi liquid up to vertex corrections, one would find
$R = \alpha T_1 T = {\mbox const.}$   This method of comparing 
temperature dependences becomes particularly powerful if
one or both of the quantities becomes strongly temperature dependent, as
would happen, for example, at the opening of a gap.   This observation
forms the basis of the remainder of our discussion of spin--charge
separated systems.

\section{Application to experiment}

\subsection{High temperature superconductors}

Perhaps the most interesting candidates for a spin--charge separated 
description are the cuprate high temperature superconductors. 
Underdoped samples exhibit an exotic metallic state with systematic unconventional
temperature dependence of transport coefficients and, for some range 
of temperatures above the superconducting transition temperature 
$T_c$, evidence of a gap or ``pseudogap'' 
opening, at least for spin excitations.   Spin--charge separated theories of 
the ``normal'' state of these systems have been advanced; in these the 
pseudogap which opens below some temperature $T^*$ is uniquely 
associated with spin excitation --- it is a spin--gap 
\cite{spingap}.   It is 
generally believed that below $T_c$ spin and charge are reunited
in Cooper pairs of d--wave symmetry with more or less free Fermionic
quasiparticle excitations in the nodes of the gap.
Alternative models of the pseudogap have been proposed which are not
spin--charge separated.   We contend that the comparison of 
XPS and NMR could help us to choose between them.

Important evidence for the opening of a pseudogap in the cuprates was
gained from the temperature dependence of the local spin 
susceptibility, as measured by NMR \cite{T*}.  
If quasiparticles in the normal 
state are electron--like, as is generally the case in non 
spin--charge separated theories, 
then the pseudogap must reflect a
general suppression of the density of states at 
the Fermi surface, and should be felt in the spin and charge 
channels.  The temperature dependence of the local charge 
susceptibility, as measured by XPS, should therefore in the first 
approximation be the same as the 
temperature dependence of the local spin susceptibility.
If, on the other hand, low energy spin and charge excitations
``decouple'' at $T*$, this may have observable consequences for the 
evolution of NMR and XPS lineshapes in the pseudogap regime.

As stressed above, we have in the interests of clarity adopted a 
deliberately naive picture of the screening response which determines XPS 
lineshapes in strongly correlated electrons systems. 
In the case of the Cu--O lattice of the cuprate superconductors, a 
hybridization of metallic valence band states (Cu $d$--electrons) with 
ligand core levels (O $p$--orbitals) --- the so called ``Zhang--Rice 
singlet'' \cite{zhang} --- is believed to be an essential part of the 
microscopic physics of these systems.
The interplay of charge transfer and strong onsite interaction 
effects underlying this hybridization can have important consequences 
for photoemission, for example in the case of Cu $2p$ states in the undoped 
cuprate LaCuO$_{3}$, where it leads to the appearance of a hierarchy of 
``satellite'' peaks alongside the principle core line in XPS spectra 
\cite{sawatzsky1}.  
Comparable interaction/charge transfer effects are believed to occur in many 
transition metal systems \cite{zaanan}.  In fact this correlated screening of 
core holes on Cu site in cluster models of the Cu--0 plane need not be local 
even in the sense of being restricted to reconfigurations of charge on the 
orbitals adjoining the ``impurity'' atom, but may extend across several unit 
cells \cite{sawatzky2}.  Since photoemission is a surface sensitive 
technique, care must also be taken to eleminate spurious extrinsic features 
which mimic satellite structure from core spectra \cite{sharma}.
   
A complete theory of lineshapes in these systems would of necessity involve
both a proper treatment of the many--orbital atomic physics of 
the Cu and 0 atoms, and the many body physics of the conduction
electrons, for each rival model of each phase of the system.   
None the less, the Doniach--Sunjic lineshape {\it has} proved useful
in fitting some core levels in the cuprates, and many of the important
low energy features of a successful many--body  theory of the cuprates 
can be expected to be independent of the detailed structure 
of the underlying microscopic model.  In particular the gross difference 
between low energy spin and charge susceptibilities should persist in any
spin--charge separated scenario, and so the comparison of XPS and NMR
should still be useful.
The lineshapes of core levels of atoms 
neighbouring the $Cu$--$O$ plane which are not hybridized strongly 
with the valence band electrons (for example the shallow $p$ levels 
of $La$ and $Sr$ in ($La_{1-x}Sr_{x})_2 Cu O_{4-y}$ \cite{fujimori})
should in general be much simpler to calculate, although experimental spectra 
can be complicated by the near degeneracy of inequivalent levels on different 
atoms.

So far as we are aware, no serious analysis has been made of
changes in core level lineshapes as a function of phase and 
temperature in the cuprates, even at the level of an attempt to extract
the temperature dependence of the asymmetry exponent $\alpha$.
However some evidence that the local spin susceptibility changes at $T^{*}$ 
is provided by the shift of XPS lines at $T^*$ in underdoped samples, which 
mimics that at $T_c$ in overdoped samples \cite{karlson}.  
A reduction of the local static charge 
susceptibility due to the opening of a gap will in general lead to a shift of 
lines to lower binding energies.  Unfortunately this shift in XPS lines
cannot be compared directly to the NMR Knight shift as the Knight 
shift is determined by the uniform and not the local spin susceptibility.
None the less, as described above, $1/T_1T$ and $\alpha$ {\it can} be 
compared directly, and the comparison of their temperature 
dependences might yield interesting information about the relative 
effect of the pseudogap in spin and charge channels.
The most interesting ranges of temperatures to study in this context 
would be the regions around $T^{*}$ at which the pseudogap opens, and $T_c$
at which spin and charge are implicitly believed to recombine.

\subsection{Quasi--one dimensional CDW systems}

It is much easier to make concrete statements about spin--charge 
separation in quasi--one dimensional materials where the relevant
theoretical models have already been widely studied.
We therefore now turn to the analysis of XPS and NMR in a 
purely one--dimensional context, which we believe to be relevant 
to the opening of a pseudogap in quasi--one dimensional CDW systems.
Our aim is to provide a useful starting point for 
comparison with experiment, not a definitive
treatment of XPS in pseudogapped systems, but
it is still important to restate the limitations 
of our perturbative approach.

The weak--coupling treatment of XPS spectra is based on the 
assumption that the lineshape is dominated by the 
powerlaw divergence with exponent $1-\alpha$ at threshold, 
and that $\alpha$ is small enough to be reliably estimated
from perturbation theory.  
Strictly this is only true at vanishing core hole 
coupling and zero temperature, but 
lineshapes based on these assumptions have been 
applied successfully to metals over a wide
range of temperatures, and so provide a reasonable starting
point for any metallic system.

We assume that at sufficiently high temperatures 
the system is not only metallic but that $\alpha$ is 
temperature independent, as in the free electron gas.
We then calculate
the dominant temperature corrections arising as the temperature of 
the system is reduced and the mechanism driving the pseudogap becomes 
effective.
A number of special features arise in one--dimension ($1D$).  
These will be dealt with in context.  

In a $1D$ free electron gas, the Fermi surface comprises
distinct left and right Fermi points and there are two important 
scattering channels corresponding to ``forward scattering'' at 
either Fermi point (momentum transfer $q\approx 0$), and ``backward 
scattering'' between Fermi points 
(momentum transfer $q\approx \pm 2k_f$). 
There is no distinction between spin and charge susceptibilities, 
and in the weak coupling limit considered above the XPS exponent is given by
\be
\label{eqn:alpha0}
\alpha &=& n_0^2 \left[ \mid V_0 \mid^2 + \mid V_{2k_f} \mid^2 \right]
\en
and the NMR relaxation time by
\be
\label{eqn:1TT0}
\frac{1}{T_1 T} &=& n_0^2 \left[ F_0 + F_{2k_f} \right]
\en

As remarked above, both $\alpha$ and the $1/T_1T$ 
are {\it temperature independent} (the Korringa law).  
These expressions become temperature 
dependent when interaction is included in one of two ways.   
If a pseudogap opens through interaction 
with an ``external field'' (\ie {\it without} destroying 
the electron--like nature of quasi--particles) it will {\it suppress}
the density of states at the Fermi energy, giving the a temperature 
dependence to the prefactor $n_0^2$ for both $\alpha$ and $1/T_1T$.
In the case of one simple model of the pseudogap observed
above $T_{3D}$ in charge density wave (CDW) systems, the Lee, Rice and Anderson 
(LRA) model \cite{lra}, the suppression of $n_0^2$ with decreasing $T$ may 
be calculated explicitly \cite{taisyii}.   While the LRA model is somewhat 
crude, its application to the 
quasi--one dimensional CDW system (TaSe$_4$)$_2$I is successful enough that we 
consider it here as a valid phenomenological alternative to the 
strongly correlated ``Luttinger Liquid'' (LL) family of models 
\cite{taisy}.

Electron--electron interaction in one dimension, on the other hand, destroys all 
Fermionic quasiparticles and lends a temperature dependence 
to $\alpha$ ($1/T_1T$) through the scaling of the $2k_f$ ``backscattering''
component of the relevant charge (spin) susceptibility.   This 
reflects the critical nature of the $1D$ interacting Fermi gas.   
The generic description of a one--dimensional Fermi liquid is the LL
\cite{tomonaga,voit}, 
a fully spin--charge separated state (according to definition 
\ref{eqn:defn}), whose excitations are long--wave length spin and charge 
density fluctuations described by the Tomonaga--Luttinger model
\be
\label{eqn:LLH}
{H}^{LL} &=& { H}_{\rho}^{LL} + { H}_{\sigma}^{LL}\\
{ H}_{\rho} &=& \frac{v_{\rho}}{2\pi} \int dx
   \left[ 
   \frac{1}{K_{\rho}} \left( \partial_x \phi_{\rho}  \right)^2
   + K_{\rho} \left( \partial_x \theta_{\rho} \right)^2   
   \right]\\
{ H}_{\sigma} &=& \frac{v_{\sigma}}{2\pi} \int dx
   \left[ 
   \frac{1}{K_{\sigma}} \left( \partial_x \phi_{\sigma} \right)^2
   + K_{\sigma} \left( \partial_x \theta_{\sigma} \right)^2
   \right]
\en
where $v_{\rho(\sigma)}$ is the charge(spin) velocity,
and $K_{\rho(\sigma)}$ parameterizes interaction.
In general, for repulsive interaction $K_{\rho} < 1$ and 
spin--rotation invariance requires $K_{\sigma} = 1$.   
The bosonic field $\phi_{\rho(\sigma)}$ and its canonical
conjugate $\partial_x \theta_{\rho(\sigma)}$ are related to the 
charge(spin) density and charge(spin) current density of physical 
electrons according to 
$\rho_{\rho(\sigma)} = -\sqrt{2/\pi}\partial_x \phi_{\rho(\sigma)} $
and $j_{\rho(\sigma)} = -\sqrt{2}K_{\rho(\sigma)}v_{\rho(\sigma)} 
\partial_x \theta_{\rho(\sigma)} $

One subtlety which must be kept in mind is that
the independent spin and charge excitations for $q \approx 0$
are mixed by processes involving a $q \approx 2k_f$ momentum 
transfer.   This is a special feature of one dimension which 
complicates but does not invalidate the type of analysis we wish to 
pursue.

For a Luttinger Liquid (Eqn. \ref{eqn:LLH}) the scaling of the XPS and NMR 
responses away from weak coupling may be described by
\be
\label{eqn:alphaLL}
\alpha^{LL} &=& n_0^2 \left[ \beta_{\rho} \mid V_0 \mid^2  + \mid V_{2k_f} \mid^2 
   \left( \frac{T}{T_0} \right)^{-\gamma} \right]\\
\label{eqn:1TTLL}
\frac{1}{T_1^{LL} T} &=& n_0^2 \left[ \beta_{\sigma }F_0  + F_{2k_f} 
    \left( \frac{T}{T_0} \right)^{-\gamma}\right]\\
\beta_{\rho(\sigma)} &=& K_{\rho(\sigma)} 
\left(\frac{v_{\rho(\sigma)}}{v_f} \right)^2 
\quad \quad \gamma = 2 - K_{\rho} - K_{\sigma} 
\en
where $T_0$ is a crossover temperature scale naively of order the bandwidth.
In the non--interacting limit ($K_{\rho} = K_{\sigma} =1$, $v_{\rho} = 
v_f$), $\gamma = 0$ and $\beta = 1$, so we recover \ref{eqn:alpha0} 
and \ref{eqn:1TT0} \cite{voit}.  
In general $\gamma > 0$ and $2k_f$ contributions to the overall response 
are not suppressed but {\it enhanced} with decreasing temperature.

The results \ref{eqn:alphaLL} and \ref{eqn:1TTLL} are {\it only} valid as 
a description of scaling away from weak coupling; in the interest of
simplicity we assume that is is sufficient to consider only this regime.
For spinless Fermions at $T=0$ it is believed that the backward scattering 
contribution to $\alpha^{LL}$ takes on the universal value $1/8$; in practise a 
transition to an ordered state at some intermediate temperature 
$T_{3D}$ may prevent this strong coupling limit from ever being reached.
Equation \ref{eqn:1TTLL} represents the usual prediction for NMR in a 
LL and has been applied with some success to data taken on the quasi 
one--dimension Bechgaard salts \cite{bourbonais}.  
Equation \ref{eqn:alphaLL} is the equivalent prediction for XPS in a LL.

The difference between the temperature dependence 
of the local spin and charge susceptibilities of a LL is quite 
subtle because it is dominated by the mixing of spin and charge 
excitations through ``backscattering'' processes. 
However the LL does not offer a good description of the 
physics of quasi--one dimensional CDW systems such as (TaSe$_4$)$_2$I, 
as these exhibit clear evidence of a pseudogap and have dominant CDW 
fluctuations.   

We can break the symmetry between CDW and SDW 
fluctuations inherent in the LL and obtain a gap to spin excitations
by introducing one further term in the Hamiltonian
\be
\label{eqn:HDelta}
{ H}_{\sigma}^{\Delta} &=& \frac{2g_{1\perp}}{(2\pi\epsilon)^2} \int dx
    \cos ( \sqrt{8}\phi_{\sigma} )
\en
where $g_{1\perp}$ is the strength of non current--conserving ``backscattering''
interactions and $\epsilon$ a short distance cuttoff.
In the special case $K_{\sigma} = 1/2$ the model 
may be solved explicitly by refermionization and it is found that the
spin sector corresponds to a set of (massive) fermions with 
gap $
\Delta_{\sigma} = 2g_{1\perp}/2\pi\epsilon
$
The charge sector remains ungapped.   It is known that the gap
persists for $K_{\sigma} \ne 1/2$, although it nolonger has this simple
form.
This new one dimensional electron liquid is known as the 
Luther Emery (LE) Liquid \cite{lutheremery}.

The opening of a spingap affects the temperature dependence of $\alpha$ 
and $1/T_1T$ in two ways.   Firstly the contribution to the local spin 
susceptibility from long--wavelength spin fluctuations becomes activated, 
and is exponentially small at temperatures low compared with the spin--gap 
scale (measured pseudogaps in $1D$ systems are usually much larger than 
relevant experimental temperatures).   Secondly, the existence of a 
spin coherence length 
$\xi_{\sigma}^{-1} \sim \Delta_{\sigma}/v_{\sigma}$
cuts off critical scaling in the spin channel
The results (\ref{eqn:alphaLL}) and  (\ref{eqn:1TTLL}) are therefore modified to
\be
\label{eqn:alphaLE}
\alpha^{LE} &=& n_0^2 \left[ 
   \beta_{\rho} \mid V_0 \mid^2  + \mid V_{2k_f} \mid^2 
   \left( \frac{T}{T_0} \right)^{-\gamma} 
   \right]\\
\label{eqn:1TTLE}
\frac{1}{T_1^{LE} T} &=& n_0^2 \left[ 
   \beta_{\sigma} F_0 e^{-\Delta_{\sigma}/T}
   + F_{2k_f}
   \left( \frac{\Delta_{\sigma}}{T_0} \right)^{-\gamma}
   \right]
\en
where $\beta_{\sigma} \sim { O}(1)$ 
and $\Delta_{\sigma} \sim v_f/\epsilon \exp(\pi v_f/g_1)$ are 
for $K_{\sigma} \ne 1/2$ undetermined coefficients.

The independence of spin and charge excitations is now
manifest in the very different temperature dependences of 
$\alpha$ and $1/T_1T$.  The ratio ${ R}(T)$, our proposed diagnostic 
for spin charge separation, is shown for each case in Table 
\ref{table1}, where the coefficients $a,b,c$ \etc as well as the 
exponent $\gamma$ and spingap $\Delta_{\sigma}$ are to be determined 
empirically.
The result for the LE liquid is clearly the ``smoking gun'' --- 
it is very hard to conceive any non spin--charge separated scenario 
in which the local spin--susceptibility could be activated while the 
local charge was not.

\begin{table}
\label{table1}
\lineup
\caption{Ratio ${ R}(T)$ for various one--dimensional electron liquids.}
\begin{indented}
\item[]
\begin{tabular}{llll} 
\br
          &  ${ R}(T)$\\ 
\mr
 \mbox{Free $e^{-}$} &  \mbox{const.}\\
 \mbox{LRA} &  \mbox{const.}     \\
 \mbox{LL}  &  $(a + bT^{-\gamma})/(c+d T^{-\gamma})$ \\
 \mbox{LE}  &  $(a^{\prime} + b^{\prime}T^{-\gamma})
                 /(c^{\prime} + d^{\prime}e^{-\Delta_{\sigma}/T} )$\\
\br
\end{tabular}
\end{indented}
\end{table}

One could repeat these arguments for a term like Equation \eref{eqn:HDelta} 
in the charge sector (Umklapp term, relevant at commensurate filling and
in Mott insulators), with the understanding that in this case a gap
opens in the charge and not the spin sector and so it is $\alpha$, not
$1/T_1T$, which displays activated behavior. 

\section{Conclusions}

XPS and NMR, as probes of local charge and spin susceptibility,
are individually very informative about the opening of pseudogaps
in strongly correlated systems.   The direct comparison 
of the two offers the possibility of determining whether a gap has opened to 
spin excitations without a gap opening to charge excitations (or 
vis--versa).   Since the opening of a gap to spin(charge) excitations without
an accompanying gap to charge(spin) excitations requires spin--charge
separation in the underlying physics, together they may be used to
probe directly for spin--charge separation.  It is anticipated that many 
of the arguments presented above with relation to the LE Liquid may be
carried over directly to assessment of the opening of a ``spin--gap''
in the pseudogap phase of the underdoped cuprates.

\ack

We wish to acknowledge stimulating and helpful conversations with Jim Allen, 
Yves Baer, Andrey Chubukov, Jeffery Clayhold, Franz Himpsel, Boldizsar Janko, Robert Joynt, 
Thilo Kopp, Volker Meden, and Qimiao Si, and support under DMR--9704972.

\end{document}